\documentclass[twocolumn,showpacs,preprintnumbers,superscriptaddress]{revtex4-1}

\usepackage{booktabs}      
\usepackage{diagbox}
\usepackage{multirow}     
\usepackage[table]{xcolor}

\usepackage{amsmath}
\usepackage{graphicx}
\usepackage{mathrsfs}
\usepackage{amssymb}
\usepackage{amsfonts}
\usepackage{amsmath}
\usepackage{CJK}
\usepackage{epsfig}
\usepackage{epstopdf}
\usepackage[colorlinks=true,linkcolor=blue,allcolors=blue,citecolor=blue]{hyperref}
\usepackage{dcolumn}
\usepackage{epsfig}
\usepackage{epstopdf}
\usepackage{graphicx}
\usepackage{ulem}
\usepackage[utf8]{inputenc}
\usepackage{amsthm}
\usepackage{bm}
\usepackage{color}
\usepackage{float}
\usepackage{geometry}
\begin{document}
\setlength{\textfloatsep}{2pt}
\setlength{\intextsep}{2pt}

\title{Detection of DC electric forces with zeptonewton sensitivity by single-ion phonon laser}
\author{Ya-Qi Wei}  \thanks{Co-first authors with equal contribution}
\affiliation{State Key Laboratory of Magnetic Resonance and Atomic and Molecular Physics, Wuhan Institute of Physics and Mathematics, Innovation Academy of Precision Measurement Science and Technology, Chinese Academy of Sciences, Wuhan 430071, China}
\affiliation{University of the Chinese Academy of Sciences, Beijing 100049, China}
\author{Ying-Zheng Wang}  \thanks{Co-first authors with equal contribution}
\affiliation{State Key Laboratory of Magnetic Resonance and Atomic and Molecular Physics, Wuhan Institute of Physics and Mathematics, Innovation Academy of Precision Measurement Science and Technology, Chinese Academy of Sciences, Wuhan 430071, China}
\affiliation{Laboratory of Quantum Science and Engineering, South China University of Technology, Guangzhou 510641, China}
\author{Zhi-Chao Liu}
\email{Corresponding author: liuzc@ciqtek.com}
\affiliation{State Key Laboratory of Magnetic Resonance and Atomic and Molecular Physics, Wuhan Institute of Physics and Mathematics, Innovation Academy of Precision Measurement Science and Technology, Chinese Academy of Sciences, Wuhan 430071, China}
\author{Tai-Hao Cui}
\affiliation{State Key Laboratory of Magnetic Resonance and Atomic and Molecular Physics, Wuhan Institute of Physics and Mathematics, Innovation Academy of Precision Measurement Science and Technology, Chinese Academy of Sciences, Wuhan 430071, China}
\affiliation{University of the Chinese Academy of Sciences, Beijing 100049, China}
\author{Liang Chen}
\email{Corresponding author: liangchen@wipm.ac.cn}
\affiliation{State Key Laboratory of Magnetic Resonance and Atomic and Molecular Physics, Wuhan Institute of Physics and Mathematics, Innovation Academy of Precision Measurement Science and Technology, Chinese Academy of Sciences, Wuhan 430071, China}
\affiliation{Research Center for Quantum Precision Measurement, Guangzhou Institute of Industry Technology, Guangzhou, 511458, China }
\author{Ji Li}
\affiliation{State Key Laboratory of Magnetic Resonance and Atomic and Molecular Physics, Wuhan Institute of Physics and Mathematics, Innovation Academy of Precision Measurement Science and Technology, Chinese Academy of Sciences, Wuhan 430071, China}
\affiliation{University of the Chinese Academy of Sciences, Beijing 100049, China}
\author{Shuang-Qin Dai}
\affiliation{State Key Laboratory of Magnetic Resonance and Atomic and Molecular Physics, Wuhan Institute of Physics and Mathematics, Innovation Academy of Precision Measurement Science and Technology, Chinese Academy of Sciences, Wuhan 430071, China}
\affiliation{University of the Chinese Academy of Sciences, Beijing 100049, China}
\author{Fei Zhou}
\affiliation{State Key Laboratory of Magnetic Resonance and Atomic and Molecular Physics, Wuhan Institute of Physics and Mathematics, Innovation Academy of Precision Measurement Science and Technology, Chinese Academy of Sciences, Wuhan 430071, China}
\affiliation{Research Center for Quantum Precision Measurement,  Guangzhou Institute of Industry Technology,  Guangzhou, 511458, China }
\author{Mang Feng}
\email{Corresponding author: mangfeng@wipm.ac.cn}
\affiliation{State Key Laboratory of Magnetic Resonance and Atomic and Molecular Physics, Wuhan Institute of Physics and Mathematics, Innovation Academy of Precision Measurement Science and Technology, Chinese Academy of Sciences, Wuhan 430071, China}
\affiliation{Research Center for Quantum Precision Measurement, Guangzhou Institute of Industry Technology, Guangzhou, 511458, China }
\affiliation{School of Physics, Zhengzhou University, Zhengzhou 450001, China}
\affiliation{Department of Physics, Zhejiang Normal University, Jinhua 321004, China}

\begin{abstract}
Detecting extremely small forces helps exploring new physics quantitatively. Here we demonstrate that the phonon laser made of a single trapped $^{40}$Ca$^{+}$ ion behaves as an exquisite sensor for small force measurement. We report our successful detection of small electric forces regarding the DC trapping potential with sensitivity of 2.41$\pm$0.49 zN/$\sqrt{{\rm Hz}}$, with the ion only under Doppler cooling, based on the injection-locking of the oscillation phase of the phonon laser in addition to the classical squeezing applied to suppress the measurement uncertainty. We anticipate that such a single-ion sensor would reach a much better force detection sensitivity in the future once the trapping system is further improved and the fluorescence collection efficiency is further enhanced. 
\end{abstract}
\maketitle

\section{Introduction}
Forces are associated with physical interaction and thus developing high-sensitivity tools of force measurement is vital to quantitative investigation of new physics. Atomic sensors developed over past years have enabled the detection of physical quantities relevant to force at a higher level of sensitivity than ever before due to their superiority of high resolution, fast response and broad range of adjustibility \cite{Hosten2016,Schreppler2019,Gilmore2017,Mamin2001,Lecocq2015,Caves1980,Degen2017,Kotler12011,Kim2017}. Studies in this aspect range from fundamental physics, detecting such as magnetic forces \cite{mag1,mag2}, electric forces \cite{sur1,sur2,sur3,Blums2018,Bollinger2021} and quantum-relevant forces \cite{quantum1,quantum2}, to potential application \cite{appli}.

Trapped atomic ions are a well-controlled quantum system characterized with high-precision manipulation and readily detectable fluorescence signal. The strong coupling to external electric fields makes the trapped ions excellent detectors of small forces due to the electric field change. Previous studies have achieved, using trapped cold ions, measurement of small forces, for instance by observing the gravity variance of the ion's displacement \cite{Knunz2010} or by the phase-coherent Doppler velocimetry \cite{Biercuk2010}. Small forces can also be detected by real-space imaging to resolve the excited micrometre-scale motional excursions \cite{motion1}. In particular, a recent detection of the sub-attonewton force was accomplished by optically measuring the ion's displacement in three dimensions based on phase Fresnel lens \cite{Blums2018}.

In the present work, we report our ultrasensitive detection of small electric forces regarding the DC trapping potential in an ion trap, which reaches the zeptonewton (zN)'s order of magnitude. The detection is achieved by a super-sensitive probe made of a single trapped $^{40}$Ca$^{+}$ ion. In our experiment, no sideband cooling of the ion is required. After Doppler cooling, the ion is excited to be the state of phonon laser, which is locked on the oscillation frequency by injection signal. Such a phonon laser, i.e., an amplitude-amplified harmonic oscillator, is quite sensitive to the external disturbance, as reported previously \cite{Knunz2010,Vahala2009,Grudinin2010,Ip2018,He2016,Hush2015,Dominguez2017,Liu2021}. As elucidated below, the DC electric force is detected due to the sensitive modification of the oscillation phase of the phonon laser, based on the method of fitting the fluorescence curve directly acquired from the photomultiplier tube (PMT) along with the synchronized measurement of the injection-locking signal \cite{Liu2021}.


Our detection method and results have the following distinct features in comparison with the previous work. Despite employment of the ion's phonon laser, which is similar to that in \cite{Knunz2010,Liu2021}, here we aim to detect the tiny force due to modification of DC trapping potential, which is very different from the detection of the alternating electric force originating from an externally applied weak rf signal \cite{Knunz2010,Liu2021}. Besides, we acquire a better sensitivity of the sensor by means of the squeezing applied on the relevant quantities to be detected, which largely reduces the measurement uncertainty with respect to that in \cite{Knunz2010,Liu2021}. More importantly, with these efforts we have finally achieved the detection sensitivity of 2.41$\pm$0.49 zN/$\sqrt{{\rm Hz}}$ for small electric forces on the ion due to the tiny change of the DC trapping potential, which is better than the detection using phase Fresnel lens \cite{Blums2018}  by two orders of magnitude.

\begin{figure*}
\includegraphics[width=12.3 cm,height=13 cm]{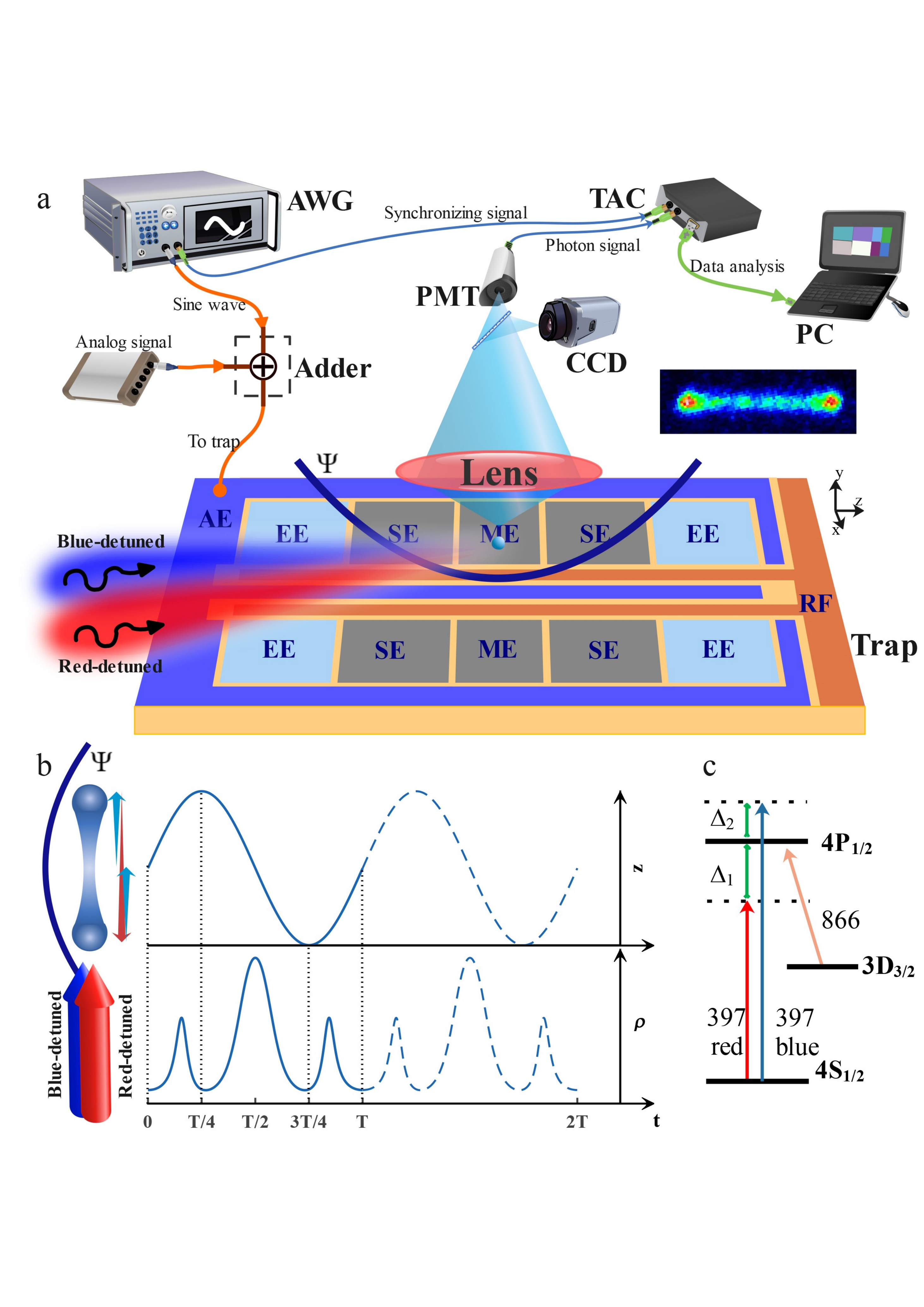}
\caption{Experimental system. (a) Sketch of the experimental setup. The $^{40}$Ca$^{+}$ ion is confined in a surface-electrode trap by the potential $\Psi(t)$ regarding a time-varying electromagnetic field. The trap is composed of four endcap electrodes (EEs), four steering electrodes (SEs), two middle electrodes (MEs), an axial electrode (AE) and a radio-frequency electrode (RF). Two laser beams address the $^{40}$Ca$^{+}$ ion, where the red-detuned laser is for Doppler cooling and the blue-detuned one is for phonon excitation. The injected signal with frequency $\omega_{i}$, for locking the ion's oscillation frequency, is applied on AE, and the squeezing signal with frequency doubling of $\omega_{i}$ is applied on EEs for reducing the thermal noise when required. The arbitrary wave generator (AWG) connected to AE and EEs is employed to provide synchronization signal for injection-locking and squeezing. The DC voltage applied on AE is
output from a homemade high-precision DC voltage source, which provides the tiny DC electric force felt by the ion. The Adder is employed to combine the DC and injection-locking signals. The photon-arrival-time transformed to voltage is measured by the time-to-amplitude converter (TAC). The ratio of the beam splitter for the PMT to the charge-coupled-device (CCD) is set to be 7/3. (b) Ion's motional track and phonon counts. The magnified image of the trapped ion oscillating over time is plotted with the corresponding sine curve for the ion's position over time (Upper panel) and the corresponding total scattering rate $\rho$, calculated by Eq. (\ref{eq3}), regarding the spontaneous emission with Doppler effect considered (Lower panel). (c) Level scheme of $^{40}$Ca$^{+}$ for our purpose, where the 397 red (blue) indicates the red (blue)-detuned beam of the 397-nm laser and 866 means the 866-nm laser for repumping. }
\label{fig1}
\end{figure*}

\section{Experimental system and scheme}
\subsection{Experimental system}
Our sensitive detection of forces is performed by an injection-locked phonon laser regarding a single ion confined in a surface-electrode trap (SET), as sketched in Fig. \ref{fig1}a. The SET employed here was introduced previously \cite{House2008,Wan2013,Brownnutt2015,Liu2020}, which is a 500-$\mu$m scale planar trap with five electrodes made of copper on a vacuum-compatible printed circuit board substrate. The radio-frequency voltage leads to a secular oscillation frequency of $\omega_{x}/2\pi = 603.40 \pm 0.10$ kHz and $\omega_{y}/2\pi = 950.30 \pm 0.10$ kHz in the radial direction, and the DC voltage applied along the axial direction produces a secular oscillation frequency of $\omega_{z}/2\pi = 165.04 \pm 0.01$ kHz.  In our experiment, the ion is confined with 800 $\mu$m above the surface of the SET. Besides, due to the much steeper potentials in $x$ and $y$ axes, we simply consider the injection-locked phonon laser regarding the axial motion of the ion.

Our phonon laser is produced by two 397-nm laser beams, stimulating the oscillation amplification of a trapped $^{40}$Ca$^{+}$ ion (Fig. \ref{fig1}b). One of the beams is red-detuned by $\Delta_{1}/2\pi$ = -70 MHz, which is for cooling the ion, and the other with blue-detuning of $\Delta_{2}/2\pi$ = 30 MHz excites the phonons. The ion is initially Doppler-cooled to the temperature about 5 mK under the cooling laser, and to have the phonon laser, we must keep the oscillation amplification stable \cite{Vahala2009}. To this end, we tune the laser beams elaborately to an appropriate intensity ratio, i.e., $I_{2}/I_{1}$ =0.25 (blue-detuned beam is weaker than the red-detuned one by 3/4), which yields the phonon laser with the oscillation amplitude approximately equal to 18 $\mu$m. Moreover, in our case, the three-level structure under a saturated 866-nm repumping laser as sketched in Fig. \ref{fig1}c, could be considered as an effective two-level system in the case that the decay from $P_{1/2}$ to $S_{1/2}$ is much larger than that from $P_{1/2}$ to $D_{3/2}$ \cite{Yan2019}. As mentioned above, the phonon laser produced is solely related to the $z-$axis motional degree of freedom of the ion due to the large frequency differences in different directions, and due to the same reason,
the injection-locking signal applied to lock the oscillation frequency of the phonon laser to $\omega_{z}$ has no influence on other two directions.

There is a low pass filter on the DC electrodes for suppressing the RF noises, which would reduce the applied injection-locking signal. But we have still observed the sufficient action of the injection-locking signal on the ion. This is due to the fact that our employed injection-locking signal is of the frequency of 165.04 kHz, beyond the bandwidth of the low pass filter.
When the injection frequency $\omega_{i}$ is tuned to the locking range, the $z-$axis oscillation frequency of the phonon laser would be fixed at $\omega_{i}$ with a very sharp bandwidth \cite{Knunz2010}. In this case, the oscillation phase $\varphi$ is locked, i.e., $d\varphi /dt =$ 0. In this context, the oscillation phase varying within the locking range can be carefully quantified by,
\begin{equation}
\varphi(\tilde{\delta})=\arcsin(-\frac{\tilde{\delta}}{\omega_{m}}),
\label{eq1}
\end{equation}
where $\tilde{\delta}=\omega_{i}-\omega_{z}$ is the frequency difference between the injected signal and the free-running oscillator, and $\omega_{m}$ represents half the locking range. In our system, $\omega_{m}=eE_{i}/2mz_{0}\omega_{z}$, depending on the injected signal amplitude $E_{i}$ (also called the signal strength corresponding to the injected signal voltage V$_{i}$), the oscillation amplitude $z_{0}$, and the charge-to-mass ratio $e/m$. Different from the previous considerations regarding the slight variation of $\omega_{i}$ \cite{Knunz2010,Liu2021}, our treatment in this work is relevant to the change of $\omega_{z}$ due to the additionally applied DC voltage, which produces a DC electric force to be detected. Since the oscillation phase $\varphi$ is very sensitive to the modification of $\omega_{z}$, our detection is made when the injection locking works well, i.e., $|\tilde{\delta}|<\omega_{m}$. As elucidated later, $\varphi$ is well locked in our experiment and $\varphi$ varies linearly with $\tilde{\delta}$, which ensures a high-precision detection.

\begin{figure}
\includegraphics[width=8.6 cm]{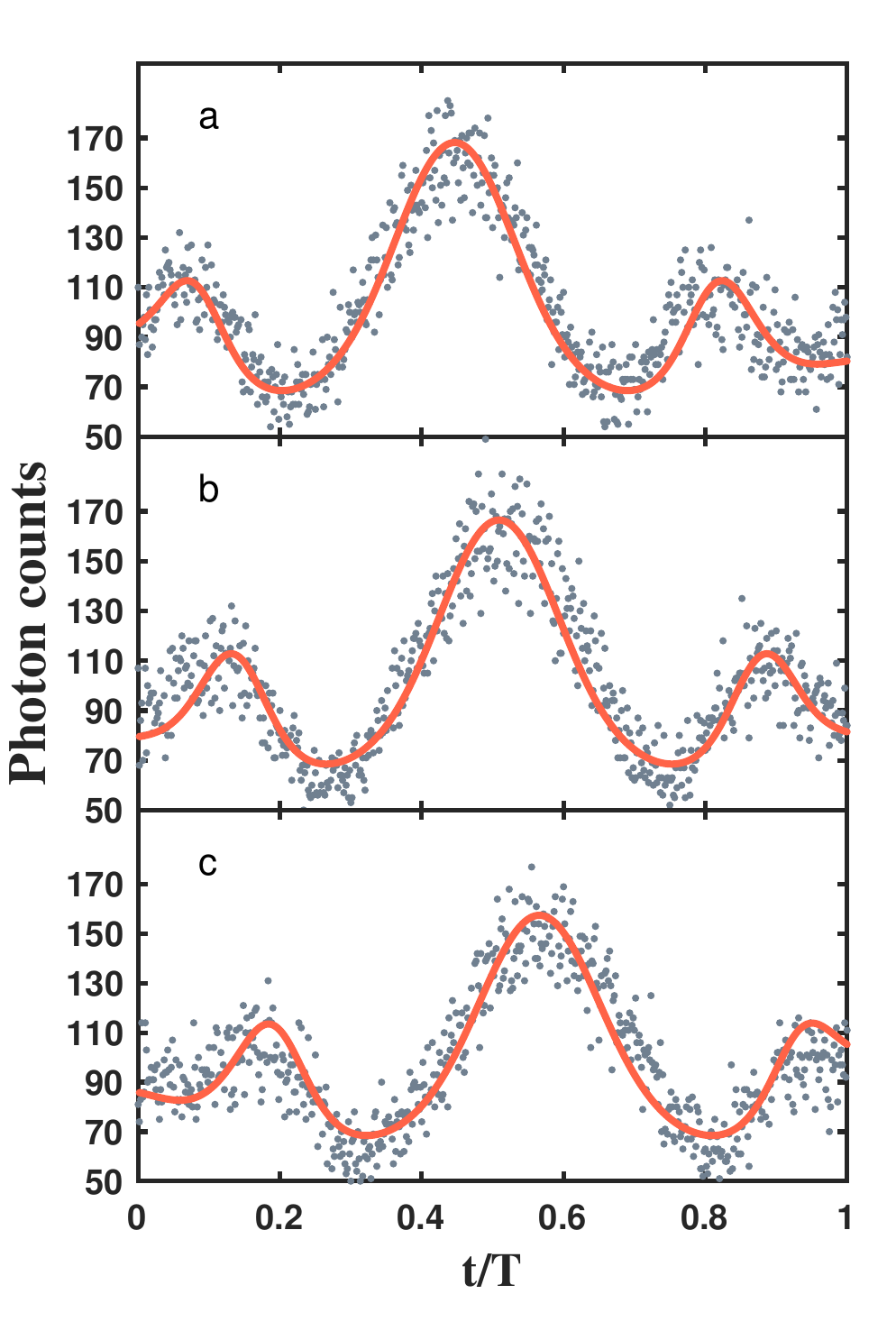}
\caption{Fitting with the accumulated photon counts for acquiring the oscillation phase of the phonon laser, where all the measurements are made at the injected signal voltage $V_{i}$= 15 mV.  (a) $z_{0}$ = 18.326 $\mu$m and $\varphi$ = 0.364 subject to the additionally applied DC voltage $V_{A}$ = -16.3 mV. (b) $z_{0}$ = 18.367 $\mu$m and $\varphi$ = -0.025 under the additionally applied DC voltage $V_{A}$ = 0 mV. (c) $z_{0}$ = 18.406 $\mu$m and $\varphi$ = -0.379 in the case of the additionally applied DC voltage $V_{A}$ =16.2 mV. The dots are experimental data and the solid curves are fitted by Eq. (\ref{eq4}). Following the optical method in Ref. \cite{Liu2021}, we set the parameter values as $\sigma_{t} = 0.8$ $\mu$s, $\alpha= 4.6\times$10$^{5}$, and $\beta= 17$.}
\label{fig2}
\end{figure}

\subsection{Optical monitoring of the phonon laser}
The phonon laser can be monitored from the CCD camera and we may observe the oscillation amplification of the phonon laser due to the locking signal applied, implying that the injection locking takes action. To detect the small forces applied on the ion, we employ the optical approach in \cite{Liu2021} to acquire the changes of the oscillation phase $\varphi$ and the amplitude $z_{0}$  of the phonon laser by PMT and TAC. The PMT records the accumulated photon counts within each measurement duration, and the TAC transforms the photon-arrival-time to voltage for graphically illustrating the curve of photon counts corresponding to the ion's oscillation (Fig. \ref{fig1}b). By fitting the experimental data regarding the photon scattering, as shown in Fig. \ref{fig2}, we may acquire the knowledge of $\varphi$ and $z_{0}$, which is much more precise than by other ways, e.g., observing the ion's motional track simply from the CCD.

For our purpose, we assume that $\varphi$ and $z_{0}$ are constants within each measurement duration $t_{m}= 10 s$ by ignoring the short-time noise. So the scattering rate $\rho_{j}$ at time $t$, under the irradiation of the 397-nm laser beams, is given by \cite{Leibfried2003},
\begin{equation}
\rho_{j}(t)=\dfrac{{\Gamma}s_{j}/(4\pi)}{1 + s_{j} + 4\left[ \frac{\Delta_{j}-k_{j}{\omega_{z}}z_{0}\cos({\omega_{z}}t + \varphi)}{\Gamma}\right]^{2}},
\label{eq3}
\end{equation}
where $\Gamma$ is the decay rate of $P_{1/2}$, $k_{j}$ is the wave vector, $\Delta_{j}$ means the red- or blue-detuning of the $j$th 397-nm beam with respect to the resonance transition (Fig. 1c), and $s_{j}$ represents the saturation parameter. $\rho_{j}(t)$ with $j=1,2$ represents the scattering rate regarding the two 397-nm laser beams with different detunings. So we have the total scattering rate $\rho(t)$ = $\rho_{1}(t)+\rho_{2}(t)$. Due to the noise involved, we need to single out information from the noise, which is carried out by the following convolution function,
\begin{equation}
P(t) = \alpha \rho(t) {\ast} G(t) + \beta,
\label{eq4}
\end{equation}
where $G(t)$ = $(1/\sqrt{2\pi}\sigma_{t})\exp[-t^{2}/(2\sigma_{t}^{2})]$ with $\sigma_{t}$ the degree of time dispersion. $\alpha$ and $\beta$, representing the factors regarding the measurement time, the fluorescence collection efficiency, the background photons, the number of unit time intervals and the signal-to-noise ratio, can be determined by fitting the experimental data, which lead to acquisition of $\varphi$ and $z_{0}$, as exemplified in Fig. \ref{fig2}. From the fitting, we find that the oscillation amplitude $z_{0}$ and phase $\varphi$ are nearly constants, consistent with our assumption. Similar to in \cite{Liu2021}, we have also observed that tuning $\varphi$ mainly changes the position of the curve in the horizontal axis and varying $z_{0}$ mainly influences the height of the second peak of the curve, indicating that $\varphi$ and $z_{0}$ are mutually independent, and thus can be acquired simultaneously by the fitting.\\

\section{Experimental implementation}
\subsection{Force sensing}
We employ the sensitivity of force detection to characterize the response of the single-ion sensor to the small force changes. Since the oscillation phase of the phonon laser is injection-locked, we define this sensitivity by the minimum detectable phase change with respect to the force $F$ within a detection bandwidth $1/\sqrt{\tau}$ , which is given by
\begin{equation}
S = \frac {\sigma_{\varphi}\sqrt{\tau}}{\frac{\partial\varphi}{\partial V_{A}}\frac{\partial V_{A}}{\partial F}},
\label{eq5}
\end{equation}
where $\sigma_{\varphi}$ is the standard deviation of the measured oscillation phase, representing the uncertainty of the data, and $\tau$ is the total measurement time equivalent to the product of the single measurement time $t_{m}$ and the measurement repetition. As the force $F$ to be detected originates from the additionally applied DC voltage $V_{A}$, we assess the detection sensitivity of the phase variation by both $\partial\varphi/\partial V_{A}$ and $\partial V_{A}/\partial F$.

\begin{figure}
\includegraphics[width=8.6 cm]{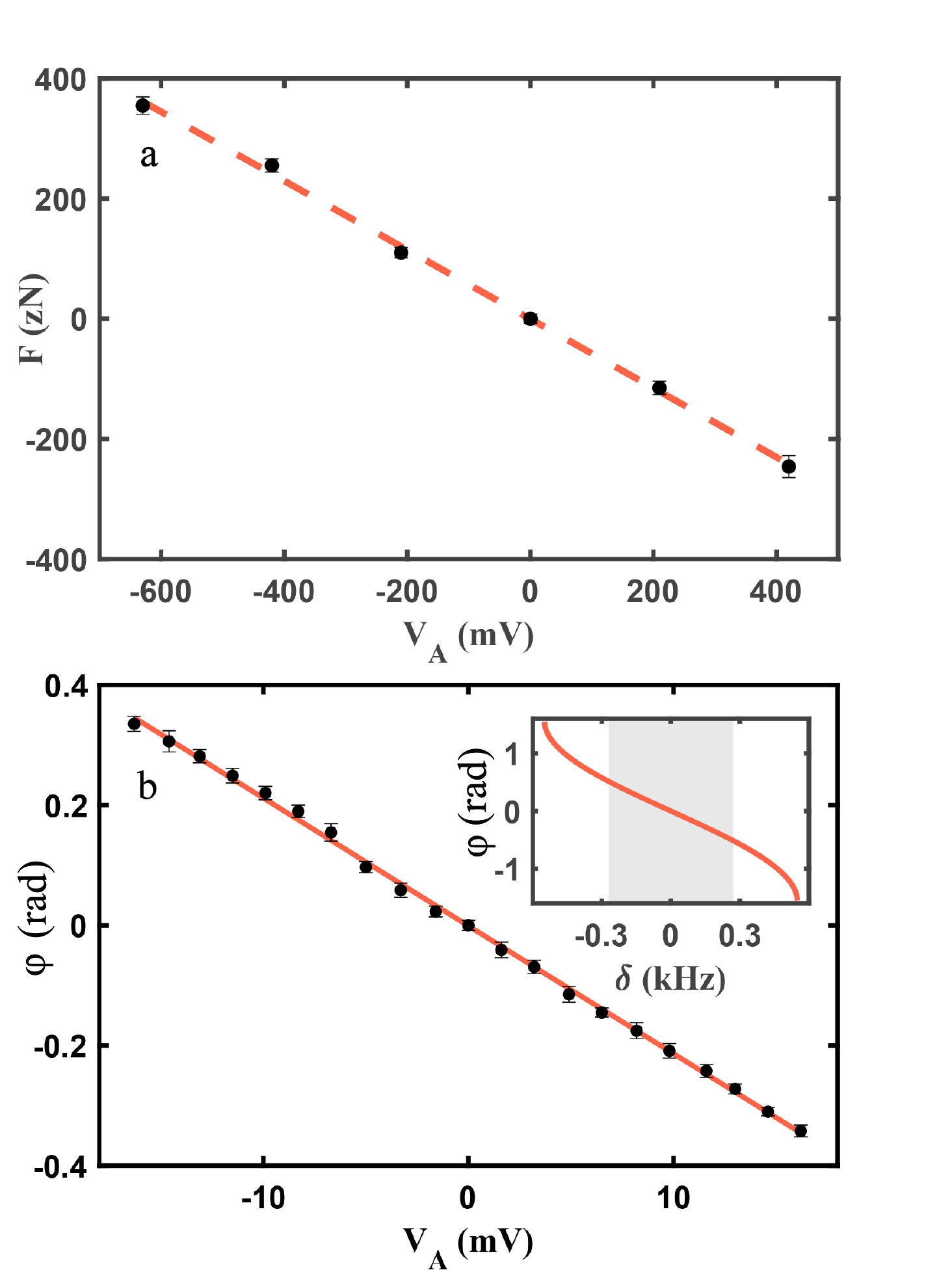}
\caption{Measurements regarding Eq. (4), where the black dots represent the experimental data and the solid or dashed line is the fitting. (a) Force varies with the additionally applied DC voltage $V_{A}$ in the absence of the injection locking. (b) The oscillation phase changes with respect to the additionally applied DC voltage $V_{A}$ subject to the injection locking, where the voltage interval of measurement is 1.6 mV. Total photon counts $N$ = 6.12$\times$10$^{4}$ are recorded by 603 unit time intervals, in which each measurement time is 10 s. Inset is the illustration of Eq. (1) with the linearly varying region marked in dark. In our experiment, since the injected signal voltage V$_{i}=$15 mV and the locking range is $2\pi\times$ 1.1 kHz, the parameters vary fully within the dark region. In both panels, the error bars represent statistical standard deviation of our experimental results by 10 measurements
for each data point. }
\label{fig3}
\end{figure}

For an oscillator system as the trapped ion, the net force to be detected due to modification of the DC trapping potential can be written as $\delta$$F = m\omega_{z}^{'2}z' - m\omega_{z}^{2}z$, in which both the trap frequency and the ion's position are changed (from $\omega_{z}$ and $z$ to $\omega_{z}^{'}$ and $z^{'}$) due to the additionally applied DC voltage. For our purpose,
we take measurements of $\partial F/\partial V_{A}$ by the ion’s displacement observed from the CCD images in the presence and absence of the large additional DC voltages on AE. These measurements are not subject to the injection locking, as shown in Fig. \ref{fig3}a, which yield ${\partial}F/{\partial}V_{A} = 575 \pm 19$ yN/mV. In contrast, the variations of the trap frequencies in three dimensions are measured by the the conventional method of the RF resonance. Then under the injection locking, we apply a small DC voltage change on the AE using a homemade high-precision DC voltage source (see Appendix for more details), which generates a tiny DC electric force on the trapped ion due to the modified trapping potential. This implies that the additionally applied DC voltage leads to variation of the oscillation phase $\varphi$ of the injection-locked phonon laser. As discussed above, $\varphi$ can be accurately measured by fitting the curve of the photon counts when the oscillation frequency is locked. By this way, we obtain a linear relationship between the DC voltage $V_{A}$ and the oscillation phase $\varphi$ with the slope ${\partial}\varphi/{\partial}V_{A} = (21.68\pm 0.47)\times 10^{-3}$ rad/mV, see Fig. \ref{fig3}b.
Due to the frequency drift, however, the measured secular frequency along $z$ axis is $\omega_{z}/2\pi = 165.04 \pm 0.01$ kHz, which yields the uncertainty of the measured phase to be 0.0127$\pm$0.0019. Therefore, the sensitivity of our force sensor in this case is 3.37$\pm$0.52 zN/$\sqrt{{\rm Hz}}$.

\begin{table}
	\caption{Comparison of the uncertainties of the measured oscillation phase $\varphi$ subject to injection locking (I) with that subject to injection locking plus additional 3 dB squeezing (II), where other parameter values are the same as in Fig. \ref{fig3}. }
\centering
	\begin{tabular}{*{3}{c}}
		\toprule
		$V_{A} (mV)$ & $\sigma_{\varphi} (\uppercase\expandafter{\romannumeral1})$ & $\sigma_{\varphi} (\uppercase\expandafter{\romannumeral2})$ \\
	  \hline
		 -16.0 & 0.0152 & 0.0090 \\
		 -14.4 & 0.0135 & 0.0097 \\
		 -12.8 & 0.0153 & 0.0073 \\
		 -11.2 & 0.0115 & 0.0087 \\
		 -9.6 & 0.0119 & 0.0118 \\
		 -8.0 & 0.0156 & 0.0092 \\
		 -6.4 & 0.0106 & 0.0084 \\
		 -4.8 & 0.0128 & 0.0094 \\
		 -3.2 & 0.0109 & 0.0060 \\
		 -1.6 & 0.0128 & 0.0078 \\
		 0.0 & 0.0092 & 0.0121 \\
		 1.6 & 0.0114 & 0.0061 \\
		 3.2 & 0.0132 & 0.0095 \\
		 4.8 & 0.0147 & 0.0118 \\
		 6.4 & 0.0116 & 0.0105 \\
		 8.0 & 0.0126 & 0.0068 \\
		 9.6 & 0.0119 & 0.0112 \\
		 11.2 & 0.0153 & 0.0086 \\
		 12.8 & 0.0131 & 0.0109 \\
		 14.4 & 0.0089 & 0.0086 \\
		 16.0 & 0.0137 & 0.0078 \\
		 mean & 0.0127 & 0.0091 \\
			\hline
		\end{tabular}
	\label{table1}
	\end{table}

\subsection{Force sensing under squeezing}
To further improve the sensitivity of our force sensor, we may consider to suppress the noise, i.e., reducing $\sigma_{\varphi}$ in our measurement of the oscillation phase using the method of classical squeezing, which is currently considered as an effective way to reduce the thermal noise in precision measurements \cite{Burd2019,Natarajan1995,Majorana1997,Briant2003,Wollman2015}.
Different from the quantum mechanical squeezing to suppress quantum noises for beating standard quantum limit, the classical squeezing reduces the effect of thermal
noise by redistributing the thermal fluctuations, yielding reduced fluctuation for the quantities to be measured. In our experiment, we apply the classical squeezing signal on the EEs, synchronizing with respect to the injection signal but with frequency doubling. So only the motion along $z$ axis could feel the squeezing effect.

Considering the deviations in the amplitude and phase, i.e., $\delta_{z}$ and $\delta_{\varphi}$, we may describe the oscillation under consideration as $z(t)$ = $[z_{0} + \delta_{z}(t)]\sin [\omega_{z}{t} + \delta_{\varphi}(t) ]$. This can be mathematically rewritten by two orthogonal components $X$ and $Y$ as \cite{Natarajan1995},
\begin{equation}
z(t) = X(t)\sin(\omega_{z}{t}) + Y(t)\cos(\omega_{z}{t}),
\label{eqs1}
\end{equation}
where, in our case, two orthogonal components can be the oscillation amplitude and phase, i.e., $X(t)$ = $z_{0}+\delta_{z}(t)$ and $Y(t)$ =  $z_{0}\delta_{\varphi}(t)$. Then we have the standard deviations of $X$ and $Y$ to be $\sigma_{X}$ = $\sigma_{z}$ and $\sigma_{Y}$ = $z_{0}\sigma_{\varphi}$. When the squeezing is applied on the system, the two orthogonal components $X$ and $Y$ vary as
\begin{equation}
\dot{X}(t)+\frac{\zeta}{2}\left(1+g\sin(2\phi)\right)X(t) = \frac{f_{x}(t)}{2\omega_{z}},
\label{eqs5}
\end{equation}
\begin{equation}
\dot{Y}(t)+\frac{\zeta}{2}\left(1-gcos(2\phi)\right)Y(t) = \frac{F_{0} + f_{y}(t)}{2\omega_{z}},
\label{eqs6}
\end{equation}
where $\zeta$ is a damping coefficient, $g$ is relevant to the trapping potential modification regarding the squeezing signal, and $F(t)$ = $F_{0}\sin(\omega_{i}{t})$ is the injection force. For our purpose, we define the random force by thermal noise as $f_{n}(t) =  f_{y}(t)\sin(\omega{t})+f_{x}(t)\cos(\omega{t})$. In addition, the second-order terms and '$\omega_{z}t$' terms are neglected in above equations \cite{Liu2021}.

Since we have injection-locked the oscillation phase $\varphi$, we may just focus on the squeezing effect on $\varphi$, i.e., the variance of $Y(t)$.
Using $\sigma_{Y}$ = $A_{0}{\sigma}_{\varphi}$, we acquire the variance of $Y(t)$ as \cite{Majorana1997,Briant2003},
\begin{equation}
\sigma_{Y}^{2}(g,\phi) = \frac{\sigma_{Y}^{2}(0)}{1 - g\cos(2\phi)},
\label{eq6}
\end{equation}
where the initial variance $\sigma_{Y}^{2}(0)= \frac{k_{B}T}{2m\omega^{2}}$ \cite{Majorana1997} means the variance without squeezing.

Squeezing the variance of the oscillation phase requires reducing $g\cos(2\phi)$, as indicated in Eq. (\ref{eq6}), which is achieved by tuning $g$ and $\phi$, respectively. To show the effect of squeezing, we have compared in Table. \ref{table1} between the uncertainties of our measured $\varphi$ in the absence and presence of the squeezing, where the 3dB squeezing occurs when $g\cos(2\phi)$ = -1. In this case, we have $\sigma_{\varphi}$ to be suppressed to 0.0091$\pm$0.0018, implying the sensitivity to be 2.41$\pm$0.49 zN/$\sqrt{{\rm Hz}}$.

\section{Discussion}
Compared to the discussion in Ref. \cite{Liu2021}, we have only presented here the detection sensitivity, but not the measurement of the smallest force. In the current work, the minimum voltage variation provided by the homemade high-precision DC voltage source is 1.6 mV, corresponding to the force change of 920 yN. But this is not the smallest force detectable by our sensor. If smaller voltage variation is available, we are able to detect smaller forces under current experimental condition. Moreover, further improvement of the sensitivity is possible, depending on enhancement of fluorescence collection efficiency, higher stability of the secular frequency, and suppression of noises.

Although the present work employs the same technique, i.e., the single-ion phonon laser, as in Ref. \cite{Liu2021}, either the forces or the quantities detected here are completely different from that in \cite{Liu2021}. The detection of the radio-frequency electric force in \cite{Liu2021} refers to the resonance absorption of the energy by the ion, while we detect here the electric force due to modification of the DC trapping potential. The former strongly influences the oscillation amplitude of the phonon laser, while the latter modifies the oscillation phase of the phonon laser. This is why we take our measurement here from the oscillation phase $\varphi$ of the phonon laser, instead of the amplitude $z_{0}$ as in \cite{Liu2021}. Meanwhile, since $\varphi$ is injection-locked, squeezing $\varphi$ is possible, which further improves the detection sensitivity by 1.39 times, as shown in Table I.

\section{Conclusion}
We have achieved experimentally an ultrasensitive detection of DC electric force by the phonon laser of a trapped ion, where both the injection-locking and squeezing applied help improving the sensitivity. In comparison with the Doppler velocimetry–based force sensing in Penning and Paul traps which are limited to narrow frequency band around the driving oscillation, our force sensing for DC electric forces is not much relevant to the trap frequency and thus with no fundamental lower frequency limit. Moreover, without using the phase Fresnel lens as in \cite{Blums2018}, our force detection, based on the ultrasensitive characteristics of the phonon laser rather than directly monitoring the ion's displacement imaging, has demonstrated a substantially higher sensitivity for very small DC electric force, which is more practical and economic in application. In particular, our detection works without the prerequisite of sideband cooling, largely reducing the experimental challenge.

We consider that the force sensing by means of detecting the phase variation of the phonon laser is more sensitive to the external disturbance than by directly detecting the ion's position deviation. In this context, we anticipate that our technique could be further optimized to achieve the force detection with higher sensitivity in the case of higher fluorescence collection efficiency, lighter ion probe and lower electric noise. One of the possible applications of our technique is to detect the very weak electrical property in some materials by moving these materials close to the vacuum chamber with the ion located inside. The other application is to detect the strength of the external electric fields. In terms of the minimum electric force 920 yN detectable in the present work, we anticipate the electric field strength as small as 57 $\mu$V/cm to be detectable with our technique. In comparison with other latest techniques, e.g., using Rydberg atoms, by which the minimum electric field strength detectable is 6.88 $\mu$V/cm \cite{photonics2022}, our detection is nearly at the same level. But considering the smaller force can be detected with our phonon laser technique as mentioned above, we are confident that our technique would provide good sensors for detecting the electric field strength weaker than 1 $\mu$V/cm. Finally, we hope that the idea of the present work would be extended to other systems for improving the techniques of quantum precision measurement \cite{SCPMA1,SCPMA2,SCPMA3,SCPMA4}.

\section*{Acknowledgments}
This work was supported by Special Project for Research and Development in Key Areas of Guangdong Province under Grant No. 2020B0303300001, by National Key Research $\&$ Development Program of China under grant No. 2017YFA0304503, and by National Natural Science Foundation of China under Grant Nos. U21A20434, 12074390, 11835011, 11734018.

\begin{appendix}
\section{DC voltage source}
The DC voltage source employed in our experiment is a homemade DC voltage micro-control circuit for controlling the AE voltage, in which the key hardware is the digital-analog-change chip AD7841 with the minimum voltage variation of 1.6 mV under our experimental setting. The system could be easily updated by replacing the digital-analog-change chip, and thus smaller voltage variations are available.

\begin{figure}[htbp]
\includegraphics[width=8.6 cm]{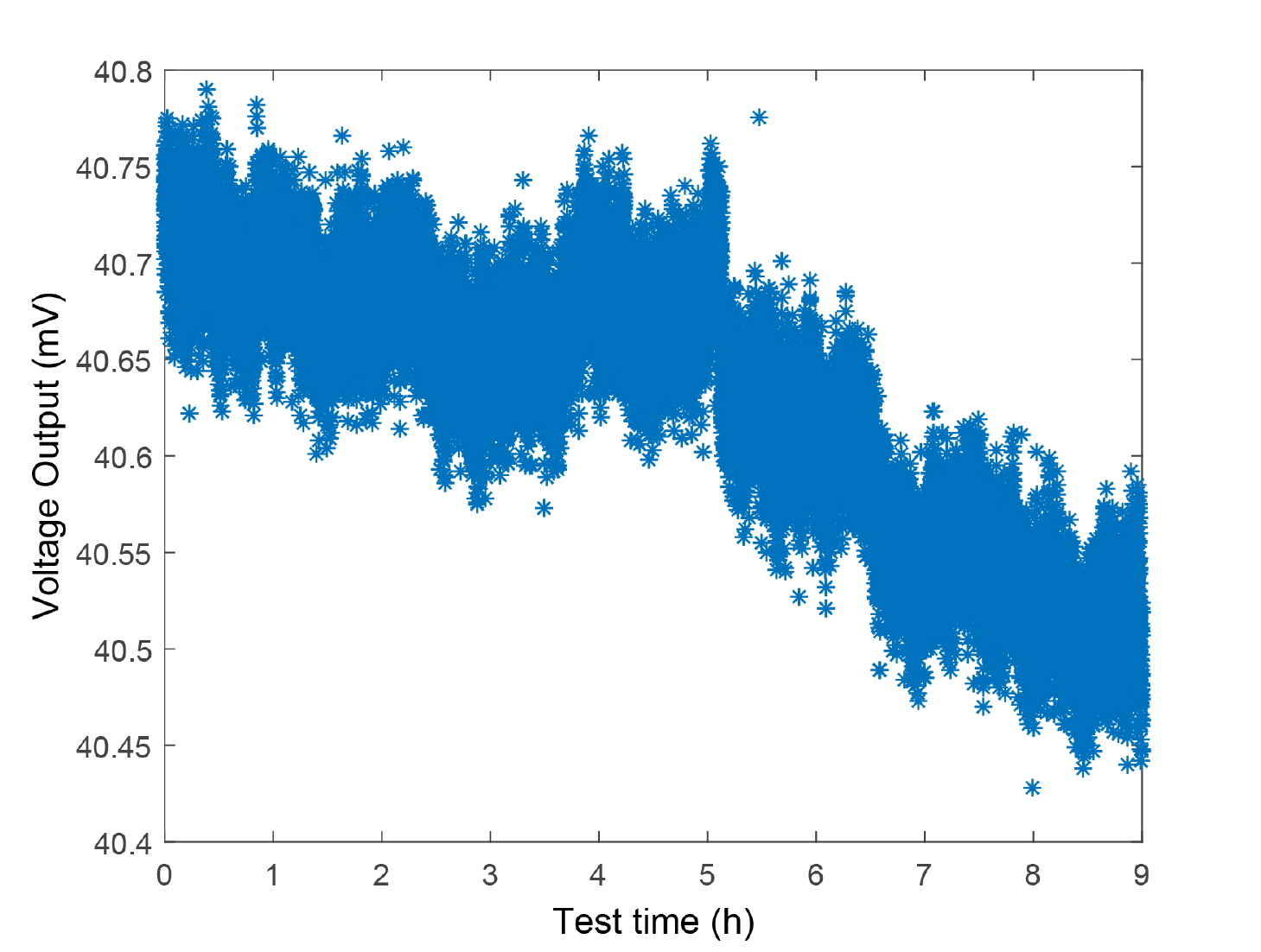}
\caption{Stability test of the DC voltage micro-control circuit for nine hours continuously with one-second sample interval, where the initial voltage output  is 40.7 mV and the last value is 40.47 mV. The largest and smallest voltage values are, respectively, 40.78 mV and 40.42 mV, implying the maximum voltage shift to be 0.36 mV.}
\label{Fig4}
\end{figure}

To ensure the accuracy of our force detection, we have to test the stability of this DC voltage micro-control circuit. To this end, we had a real-time monitoring of the DC voltage output of this circuit throughout the experiment by a 6$\frac{1}{2}$ multimeter. As shown in Fig. 4, we have monitored a tiny shift, i.e., 0.04 mV per hour, of the output voltage with respect to time, which is negligible compared to the variation of 1.6 mV in each of our experimental measurements that takes nearly one hour.
\end{appendix}


\end{document}